\documentclass[%
reprint,
superscriptaddress,
 amsmath,
 amssymb,
 aps,
prb,
floatfix,
]{revtex4-2}

\usepackage{float}
\usepackage{braket}
\usepackage{graphicx}
\usepackage[caption=false]{subfig}
\usepackage{dcolumn}
\usepackage{bm}
\usepackage{inputenc}
\usepackage{blindtext}
\usepackage{color}
\usepackage{todonotes}
\usepackage{booktabs,siunitx}
\usepackage{comment}
\usepackage{hyperref}
\hypersetup{colorlinks,citecolor=blue}

\graphicspath{{Figures/}}
\newcolumntype{~}{>{\global\let\currentrowstyle\relax}}
\newcolumntype{^}{>{\currentrowstyle}}

\DeclareMathAlphabet\mathbfcal{OMS}{cmsy}{b}{n}

\begin{document}

\newcommand{\JPcomment} [1] 
{\todo[inline,backgroundcolor=green,size=\small ,bordercolor=white]{{\bf JP:} #1}}

\preprint{APS/123-QED}

\title{Thermal Conductivity Of Monolayer Hexagonal Boron Nitride: Four-Phonon Scattering And Quantum Sampling Effects}


\author{Jos\'e Pedro Alvarinhas Batista}
\email{jpabatista@uliege.be}
\affiliation{Nanomat group, Q-MAT center, CESAM research unit and European Theoretical Spectroscopy Facility, Université de Liège, allée du 6 août, 19, B-4000 Liège, Belgium}

\author{Matthieu J. Verstraete}

\affiliation{Nanomat group, Q-MAT center, CESAM research unit and European Theoretical Spectroscopy Facility, Université de Liège, allée du 6 août, 19, B-4000 Liège, Belgium}
\affiliation{ITP, Physics Department, Utrecht University 3508 TA Utrecht, The Netherlands}

\author{Alo\"is Castellano}
\affiliation{Nanomat group, Q-MAT center, CESAM research unit and European Theoretical Spectroscopy Facility, Université de Liège, allée du 6 août, 19, B-4000 Liège, Belgium}



\begin{abstract}

Monolayer hexagonal boron nitride is a prototypical planar 2-dimensional system material and has been the subject of many investigations of its exceptional vibrational, spectroscopic and transport properties. 
The lattice thermal conductivity remains quite uncertain, with theoretical and experimental reports varying between 218 and 1060 Wm$^{-1}$K$^{-1}$. It has a strong temperature evolution and is sensitive to strain effects and isotope concentrations. While the impact of isotope scattering has been widely studied and is well understood, nuclear quantum effects and 4-phonon scattering have so far been neglected. Monolayer hexagonal boron nitride is composed of light elements, and further has its 3-phonon scattering phase space restricted by mirror plane symmetry, so these effects may be of similar order as isotope scattering, and would lead to a completely different understanding of the fundamental processes limiting the lattice thermal conductivity for this system. In this work, we use both classical and path-integral molecular dynamics, in conjunction with the Temperature Dependent Effective Potential method, to compute temperature-dependent renormalized phonons including isotope scattering, 3-phonon scattering, 4-phonon scattering and nuclear quantum effects. We show the impact of the latter two on the lattice thermal conductivity for a large temperature range, as well as their impact on the phonon lifetimes. 
Overall, our work provides a robust framework for calculations of the lattice thermal conductivity in solids, providing quantitative improvements and physical understanding that help explain the variety of results found in the literature.



\end{abstract}

                          
\maketitle

\section{Introduction}

With the rise of energetic efficiency as a worldwide priority in recent years, 
new solutions for thermal management are being sought at all scales from macro to nano. 
The strongly anisotropic and tunable nature of 2-dimensional materials suggests a broad range of applications in this domain. 
In particular, systems with very large in-plane thermal conductivity ($\kappa$) have been highly sought after for the flexibility they provide in directed cooling of semiconductor microchips.

For over a decade, simulations of thermal conductivity for layered and 2D materials pointed to graphene and monolayer hexagonal boron nitride (h-BN) as the best candidates for thermal management applications \cite{Lindsay2010, Lindsay2014, Alofi2013, Fugallo2014, Yuan2019, Lindsay2011, Sevik2011, Wu2020}. Experimental work has mostly supported this idea \cite{Cai2019, Ying2019, Balandin2008, Lee2011, Xu2014, Nika2012}, though results found in the literature can vary widely. The case of graphene is particularly extreme variance, though most studies report very high thermal conductivities.

These results have been, however, called into question more recently. Graphene's thermal conductivity has been shown to actually be lower than that of diamond if 4-phonon scattering effects are taken into account \cite{Han2023,Feng2018,Yukai2023}. In the particular case of \emph{strictly planar} 2D systems, the acoustic flexural mode was previously thought to be the main driver for the thermal conductivity. Its scattering phase space is suppressed due to mirror plane symmetry, which prevents scattering events containing an odd number of such modes \cite{Lindsay2010, Mariani2008}. The result of this suppression is that, despite the fact that the third-order interatomic force constants (IFCs) are still larger in amplitude than the fourth-order ones, the scattering phase space they span is so reduced that the sheer number of allowed 4-phonon interactions makes them dominant.

Another question that is raised for both graphene and h-BN systems is the role played by nuclear quantum effects (NQEs) in their thermal conductivities. Both are composed of light elements and have rather large Debye temperatures ($\sim$ 1680K for h-BN \cite{Ordejon2017} and $\sim$ 1800K for graphene \cite{Xie2015}). One standard approach to include NQEs in phonon simulations is the self-consistent harmonic approximation (SCHA), most commonly used in its stochastic form (sSCHA) \cite{Errea2013, Errea2014, Bianco2017, Monacelli2018, Monacelli2021}, a non-perturbative, self-consistent anharmonic phonon theory. SCHA has an effective harmonic Hamiltonian as its basis, enforcing a Gaussian distribution whose mean and variance are temperature dependent. This restriction on the distribution means that the SCHA is usually very reliable for lower temperatures, where the Gaussian approximation is often appropriate, but can fail at higher temperatures or near phase transitions, where anharmonicity can increase drastically to the point where the phonon distribution takes a more complex form, and requires extra parameters like skewness and kurtosis. A solution to this problem is to instead employ an approach based on path-integral molecular dynamics (PIMD) \cite{Craig2004, Braams2006, Morresi2021} for sampling purposes: the MD will guarantee the correct canonical distribution, and the path integral is able to handle quantum effects at low T.
Once sampled, the temperature dependent effective potential (TDEP) method \cite{Hellman2011, Hellman2013a, Hellman2013b, Knoop2024} can extract the anharmonic phonons from the dynamics. Since TDEP works by fitting IFCs directly from the snapshots it is provided with, if these come directly from MD the displacement distribution does not follow a pre-established functional form, and will remain accurate even in very anharmonic scenarios. It is worth noting that TDEP is not restricted to fitting IFCs from MD simulations, and can be used in its stochastic form as well (known as sTDEP) \cite{Shulumba2017} as another variation of the SCHA.

The PIMD + TDEP approach has been applied to metallic hydrogen \cite{Geng2022} and silicon \cite{Folkner2024}, and we gave a formal treatment and justification in a previous paper \cite{Castellano2023}, based on the Mori-Zwanzig projection operator formalism and the mode coupling approximation. The fcc phase of \textsuperscript{4}He was chosen as a test case. We showed that the method provides remarkable agreement with inelastic neutron scattering experiments and how, even at low temperatures, the PIMD based approach differs from the SCHA when comparing 3rd order IFCs. This difference can prove essential when dealing with thermal conductivity calculations, where higher order IFCs are the crucial ingredients to obtain scattering amplitudes. Combined with the fact that at high temperature the PIMD sampling based results should match their classical MD counterparts, this unified approach allows us to not only directly investigate the importance of NQEs, but also study a wide temperature range using a single consistent method.

In this work, we use TDEP combined with both classical MD and PIMD, to study the impact of NQEs and 4-phonon scattering
effects in the lattice thermal conductivity of monolayer hexagonal boron nitride. A machine learning interatomic potential (MLIP) based on the moment tensor potential (MTP) descriptor \cite{Shapeev2016, Novikov2021} is used to accelerate the sampling process for both cases. The main results of this work are separated in 3 parts. In the first part, we demonstrate the accuracy of the MLIP and address the convergence of the lattice parameter and its temperature dependence, as well as how the system's pair distribution function changes with temperature in the classical and quantum cases. In the second part, we show the consequences of the NQEs on the TDEP fit, as well as in the temperature-dependent phonon band structures. Finally, in the third part we show our main result: how the thermal conductivity evolves with temperature when including classical MD or PIMD 
and either 3 or 3+4 phonon scattering. We show that the thermal conductivity is drastically overestimated using just 3-phonon interactions, highlighting the importance of 4-phonon effects in planar 2D materials. 
The restriction imposed by mirror-plane symmetry on the acoustic flexural phonons and their interactions is such that 4-phonon scattering is required for an accurate description of the thermal conductivity. This implies that both levels of theory are essential for a proper physical understanding of the phonon-phonon interactions in monolayer h-BN, calling into question previous results for systems with similar scattering constraints.
%
%
We further show that, despite the very light masses, effects of NQEs are small in h-BN, and a classical approach therefore suffices for this system, provided full anharmonicity is taken into account.

\section{Methods}\label{sec:methods}

To build temperature dependent IFCs we rely on the TDEP method \cite{Hellman2011, Hellman2013a, Hellman2013b, Knoop2024}, which has been proven to provide a solution to the lattice dynamics problem, both in the classical and quantum cases\cite{Castellano2023} in the context of mode-coupling theory. In this approach, temperature-dependent second-order IFCs are computed from representative configurations of the canonical ensemble through a least-squares fit of the atoms' forces and displacements. Higher-order IFCs are built in succession with the residual forces, subtracting the previous order fits.

Sampling of the canonical ensemble is performed using classical and PI-MD driven by machine learning (ML). An order 22 MTP descriptor with a radial cutoff of 5.5 \AA~ is employed as per Ref.~\cite{Novikov2021}. Long range electrostatic effects are not accounted for explicitly in the MLIP, but we do not expect a significant effect on the thermal conductivity due to polar effects. In our calculations the LO and TO modes combined contribute less than 1\% to the total thermal conductivity. The training of the MLIP is done using the machine learning assisted canonical sampling (MLACS) approach \cite{MLACS2022,Mlacs2024} with PIMD. After each snapshot is added to the training set, the temperature of a Langevin thermostat is changed to a random value between 100K and 1200K. This allows us to train a single potential usable for all cases of interest. Density functional theory (DFT) calculations are done with the Abinit code \cite{Abinit1,Abinit2}, using the PBE exchange-correlation functional \cite{PBE} as well as D3 van der Waals corrections according to Ref.~\cite{Grimme2010}. A plane wave energy cutoff of 40 Ha and a 2x2x1 k-point grid were sufficient to ensure convergence for our supercell (6x6x1 with 72 atoms) total energies to a 1 meV per atom accuracy.

Molecular dynamics simulations were performed using the LAMMPS software \cite{LAMMPS} with a 0.5 fs time-step. The GJF integrator \cite{GJF} is used for the Langevin dynamics in the classical case, while the BAOAB integrator is used instead for the path-integral cases. The number of beads used in the PIMD simulations is determined by first converging the energy of the lowest temperature case (150K), and then maintaining N$_{beads} \times$T constant. The converged value is found to be N$_{beads} = 80$, implying N$_{beads} \times T = 12000$ is imposed for all temperatures.

After convergence of the MLIP, NPT simulations are launched for temperatures between 150K and 1050K, with increasing steps of 150K, in order to obtain the equilibrium volumes of the supercells. The equilibrated cells are then used for sampling under NVT conditions to obtain representative configurations of the canonical ensemble for each temperature. In the PIMD case, an extra step is performed in which the beads are used to construct a centroid trajectory where the quantum effects are implicitly included. The forces from the potential energy are averaged in the same way as the positions to obtain the centroid force (excluding the stochastic Langevin components of the force for each bead). This centroid allows us to build an approximation to the Kubo correlation function, the central object behind ensemble quantum effects in the TDEP method \cite{Castellano2023}. Finally, these configurations' displacements and forces are used in the TDEP method for the least-squares fits to obtain temperature-dependent IFCs. For the TDEP fits, the second-order IFC cutoff uses the full supercell, the third-order IFC cutoff is 5.1 \AA~ and the fourth-order IFC cutoff is 2.51 \AA.

Lattice thermal conductivities are computed using the TDEP software's in-built solver for the iterative Boltzmann transport equation (IBTE) \cite{Omini1996, Fugallo2013, Broido2007, Hellman2014}. The latest version of the solver is used as per \cite{Castellano2025}, with the full grid being employed in every case. Isotopic effects are included in the scattering processes according to the Tamura model \cite{Tamura1983, Protik2024} with a natural distribution of boron and nitrogen isotopes. These effects have been shown to be particularly important in h-BN, with substantial changes being seen in its vibrational \cite{Yuan2019}, excitonic \cite{Janzen2023} and electronic \cite{Vuong2017} properties. q-point grids of up to 360x360x1 for 3-phonon scattering cases and 64x64x1 for 4-phonon scattering cases are used. The lattice thermal conductivity is then obtained for the infinitely dense q-point grid as the y-intersect of the linear extrapolation of the $\kappa$ vs $1/q$ curve in the 3-phonon scattering cases. For 4-phonon scattering we use the $\kappa$ obtained for the largest q-grid (64x64x1) as the extrapolation can be noisy in the lowest temperature cases. An effective thickness is used for each temperature such that the $c/a$ ratio of 1.317 from bulk h-BN holds as in Ref.~\cite{Cepellotti2015}.

Finally, it is important to mention that no extra corrections have to be included due to the usage of MLIPs. It was reported in recent works \cite{Wu2024} that an extra scattering channel is introduced when calculating the lattice thermal conductivity directly from MLIP-driven MD. This problem does not appear when using MLMD + TDEP to compute phonon properties. The reason behind this is simple: the TDEP method computes the IFCs from a linear least-squares fit of the forces and their residuals into increasing displacements. As a result, the addition of a gaussian error to the forces does not change the results of the fit, as the least-squares is the maximum likelihood estimator in this case \cite{Carter_Hill2024-hq}. This constitutes an advantage to computing the lattice thermal conductivity from a quasiparticle (in this case TDEP) approach instead of directly from the MD, as the correction due to the MLIPs is expensive and complex in the latter case (involving extrapolations and noise estimators) but unnecessary in the former. It is worth noting that the error does not need to be strictly Gaussian in order for the TDEP least squares fit to remain unchanged, as any distribution with zero mean and constant variance will not affect the peak of the force distribution.

\section{Results}

\subsection{Sampling the canonical ensemble} \label{sec:sampling}

Our first goal is to establish that the ML driven MD accurately reproduces the true dynamics of our system. In particular, we want to ensure that we capture the correct temperature evolution of its lattice parameter, as the system's volume impacts its thermal conductivity, both directly through a $1/V$ factor, and indirectly through the IFCs.

\begin{figure*}
    \centering
    \includegraphics[width=0.8\textwidth]{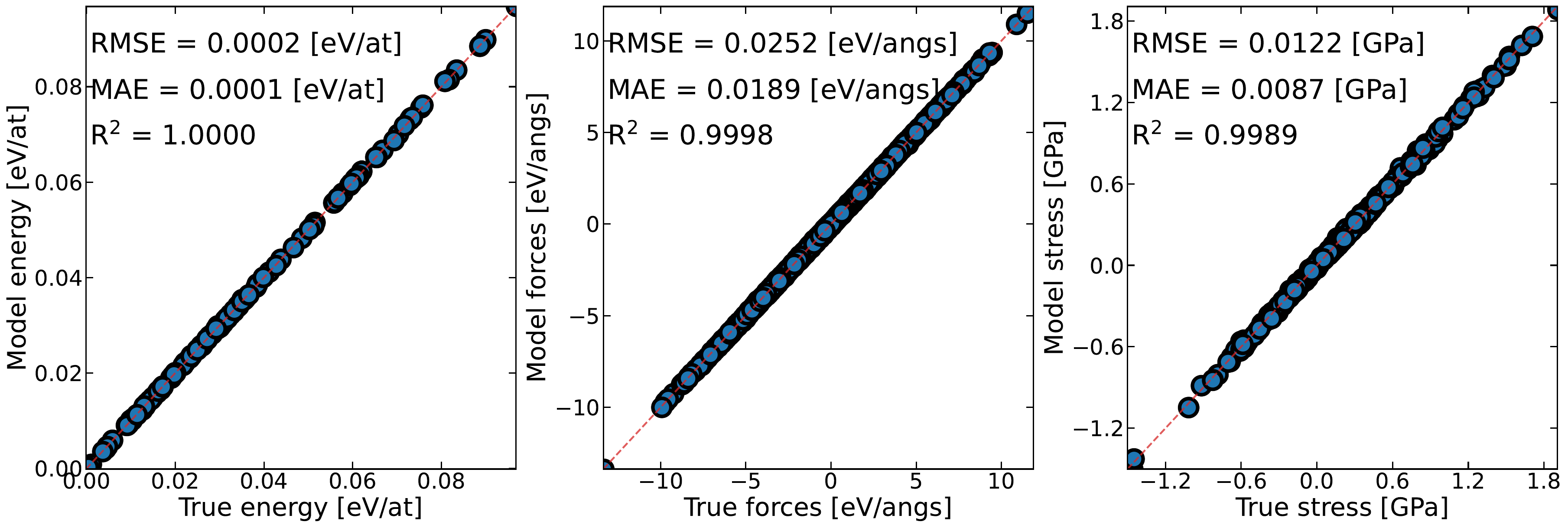}
    \caption{Correlation plots for training and test sets. The x-axis pertains to the energies, forces and stresses calculated with DFT, while the y-axis pertains to those calculated with our MLIP. The dashed red-line is a visual aid to help track the data points where these two match. The agreement is excellent as quantified by the R$^2$ coefficients in the legends.}
    \label{fig:MLIP_correlations}
\end{figure*}

We start by analyzing the quality of our MLIP. Our main goal is to guarantee that the forces reproduced by the MLIP are as close as possible to DFT, since they will be the key quantity to obtain the force constants within the TDEP approach, as well as ensuring an accurate time evolution of our system using molecular dynamics. The comparison between the energies, forces and stresses calculated with DFT and with our MLIP is shown in Fig.\ref{fig:MLIP_correlations}.

Having a highly accurate MLIP in hand, we compute the system's temperature-dependent unit cell in a range from 150K to 1050K, in steps of 150K, both in the classical and quantum cases. In thermal equilibrium, the lattice parameter $a$ for each of these unit cells is computed as the phase space average $\langle a \rangle = Tr[a \rho]$, with $\rho$ the probability distribution function given by

\begin{equation}
    \rho = \frac{e^{-\beta H}}{Z},
    \label{eq:prob density function}
\end{equation}
where $Z$ is the partition function, $H$ the Hamiltonian and $\beta = \frac{1}{k_B T}$. 
Calculating the lattice parameter directly in this way is not realistic except for very simple cases. However, using the ergodic principle one can replace the phase space average by a discrete time average in the limit of infinite number of time-steps:

\begin{equation}
    \langle a \rangle = \lim_{N\to\infty} \frac{1}{N} \sum^{N}_{t=1} a(t)
\end{equation}
This is done in practice by performing MD simulations under NPT constraints at zero pressure and averaging out the resulting lattice parameters of the snapshots of each simulation for each of the temperatures under consideration.

For cases where the number of atoms is very large, ergodicity can be considered as an innate characteristic of the system's dynamics. In our case, however, the system has 200 atoms only, which opens up the possibility that not all position and momentum degrees of freedom are explored. It can also happen that our system becomes mode-locked, and therefore constrained to exploring a closed path in phase space. This can be circumvented by initiating several different MD simulations (NPT or NVT conditions depending on what we want to calculate) for each temperature with different initial conditions, and then averaging the resulting snapshots for all of them. A large enough number of MD simulations with different starting configurations will span the full momentum phase space more efficiently, and therefore this number is a parameter to converge. Furthermore, the number of snapshots included in every MD instance also needs to be converged, as it dictates how well the phase space of that particular instance is sampled.

Together, the number of MD instances and their number of configurations control the shape and quality of our coverage of the phase space. The convergence study of these parameters is conducted directly through the temperature dependency of the lattice parameter in both classical and quantum cases. We find that convergence is achieved with 21 NPT instances with 1900 snapshots each in the classical case, and with 10 NPT instances with $\sim$ 3000
snapshots in the quantum case. These very large numbers of snapshots, and in particular the need for multiple NPT instances, are a necessity we have only observed in planar 2D systems. This is the case in h-BN, and due to the presence of the very low frequency acoustic flexural mode and very high frequency optical vibrational modes, leading to the need of both small time steps and large simulation times. These are reflected in the unusually large sample size required for an accurate lattice parameter. The lattice parameters are then fit to a cubic polynomial function of temperature to smooth them and reduce the random noise arising from molecular dynamics. Comparing our results to experimental data in \cite{Kriegel2023}, we see that the larger changes in the lattice parameter $a$ obtained with PIMD at higher temperatures are observed, indicating a need for NQEs to be included in MD simulations looking to accurately capture this observable's temperature evolution.  

\begin{figure}
    \centering
    \includegraphics[width=0.9\columnwidth]{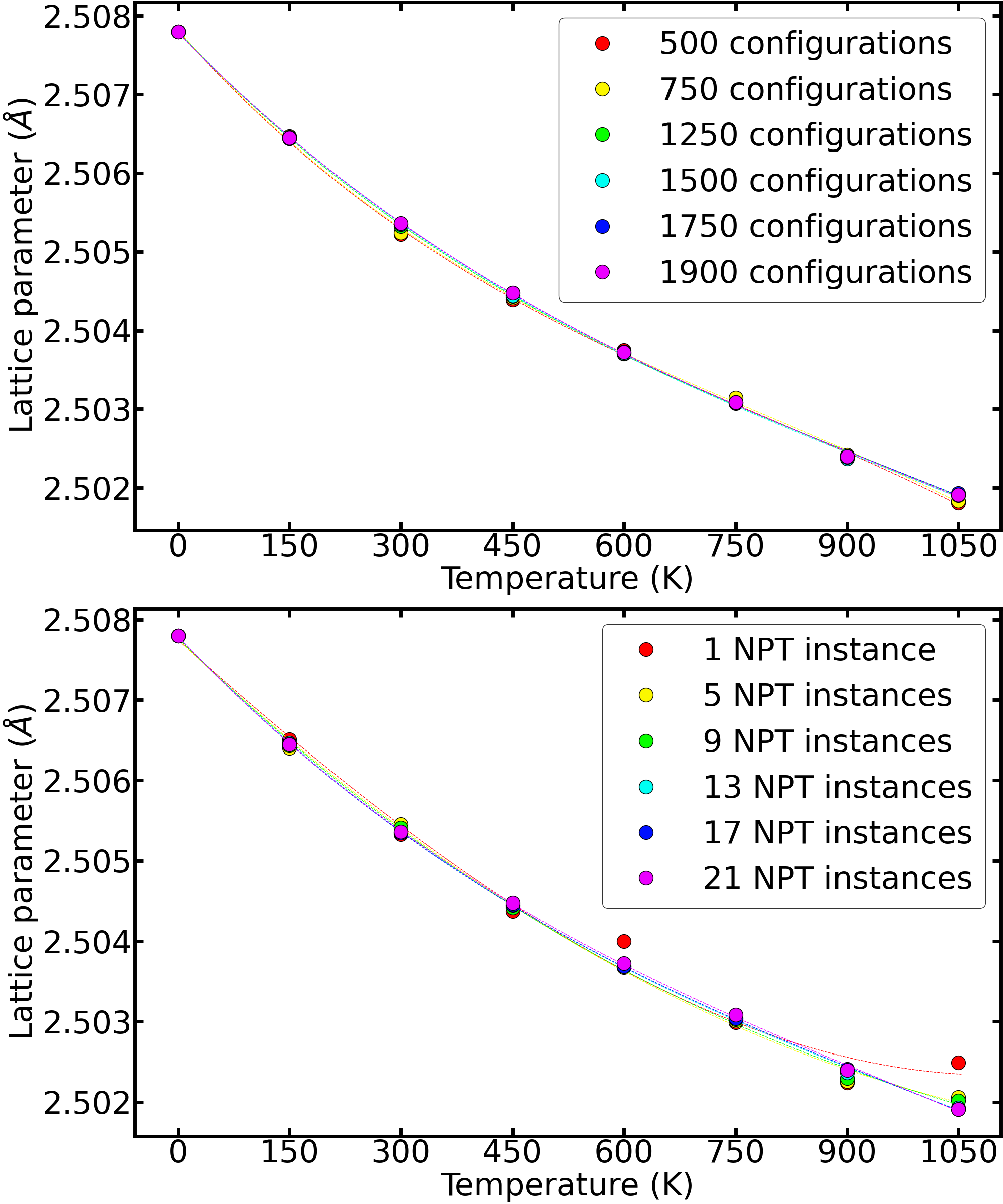}    
   
    \caption{Convergence of the temperature evolution of the lattice parameter $a$ with respect to the number of NPT instances and snapshots for the classical MD case. The number of NPT instances is fixed at 21 for the convergence of the number of snapshots. The number of snapshots is fixed at 1900 for the convergence of NPT instances. The 0 K data point comes from the full relaxation of the system driven by the MLIP, and is not used in the cubic interpolation.}
    \label{fig:lattice_param_conv_cl}
\end{figure}

\begin{figure}
    \centering
    \includegraphics[width=0.9\columnwidth]{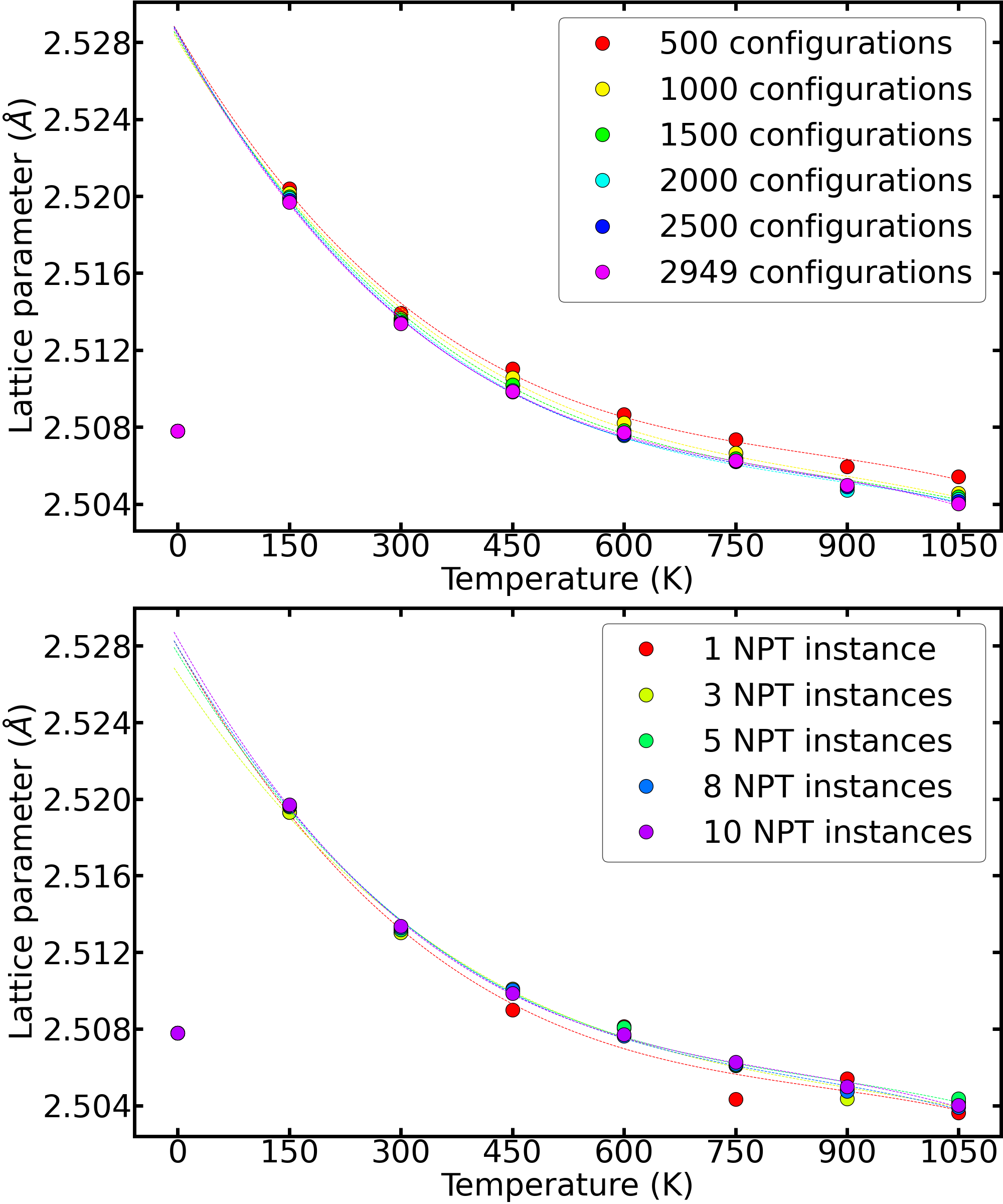}

    \caption{Convergence of the temperature evolution of the lattice parameter $a$ with respect to the number of NPT instances and snapshots for the PIMD case. The number of NPT instances is fixed at 10 for the convergence of the number of snapshots. The number of snapshots is fixed at 2949 for the convergence of NPT instances. The 0 K data point comes from the relaxation of the system driven by the MLIP and is not used in the cubic interpolation. Its deviation relative to the PIMD data shows the importance of zero-point motion in monolayer h-BN.}
    \label{fig:lattice_param_conv_pimd}
\end{figure}

At this point, it is important to verify the need for using molecular dynamics versus stochastic sampling, and whether quantum effects are visible in the sampling to the extent that the path integral approach could be required. A key quantity that enables this analysis is the pair distribution function, which contains information on the distances between pairs of atoms in the simulation supercell. If this function's shape is Gaussian, a stochastic method for sampling (e.g., the sSCHA or sTDEP) could be used without compromising accuracy; however, if it deviates substantially from a Gaussian, using molecular dynamics-based approaches becomes essential to capture the full anharmonicity of the system. Furthermore, one can compare the pair distribution functions arising from classical and path integral molecular dynamics simulations, to directly observe the impact of quantum effects and determine at what temperatures they no longer affect the distribution.

\begin{figure}
    \centering
    \includegraphics[width=\columnwidth]{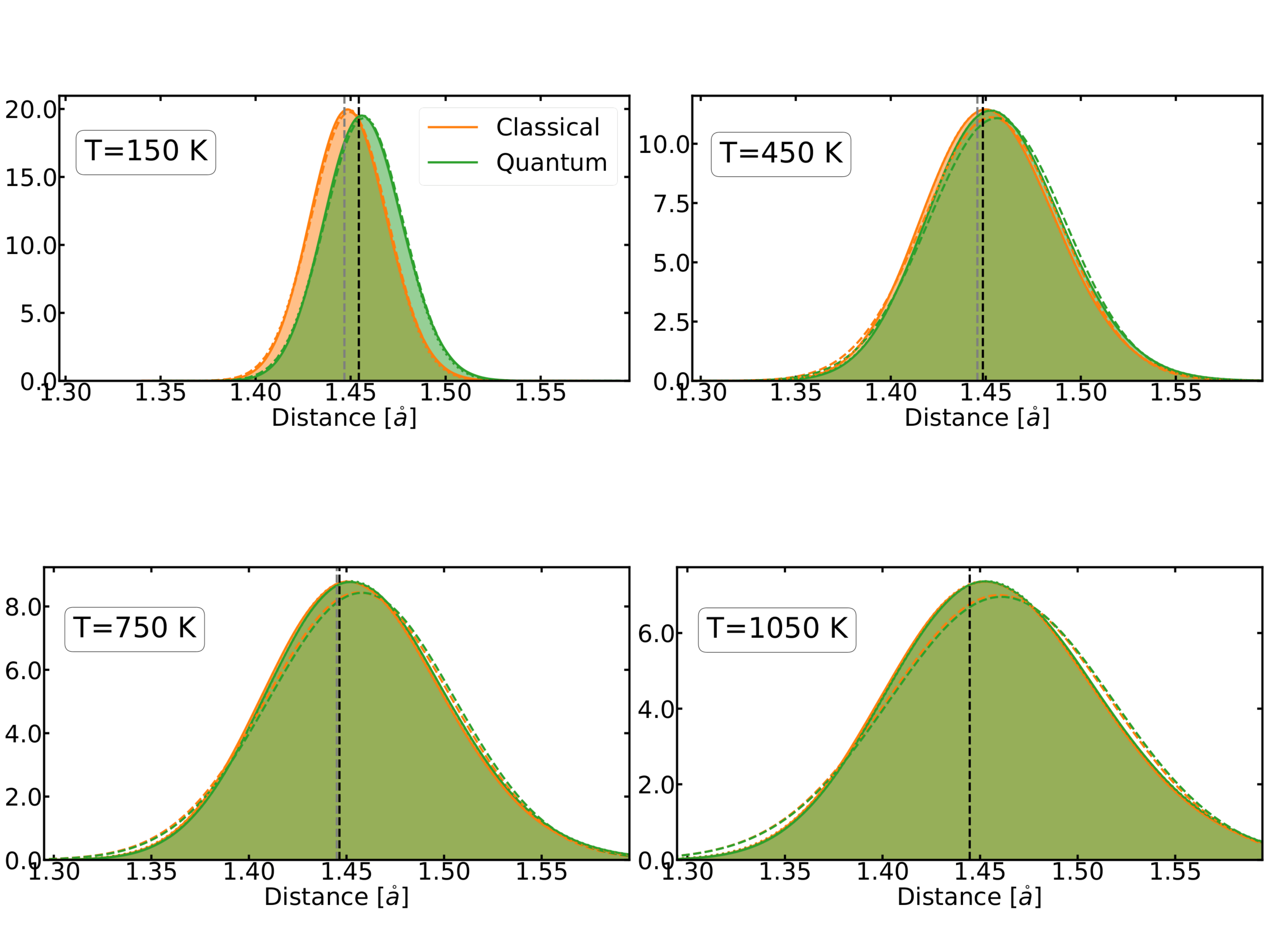}
    
    \caption{Comparison between pair-distribution functions using classical and path-integral molecular dynamics. Dashed lines correspond to Gaussian fits, and dotted lines to skewed Gaussian fits. As temperature is increased the distributions deviate from Gaussians and approach each other.}
    \label{fig:pdf_comparison}
\end{figure}

In the case of monolayer h-BN, our simulations show that at the lowest temperature the classical and quantum distributions deviate in their means, but as temperature increases, they converge, becoming very similar at room temperature. At higher temperatures, the two distributions are indistinguishable, indicating that quantum sampling effects are negligible for the vibrational properties of the system. At low temperatures the distributions are well described by Gaussians (dashed lines in Fig. \ref{fig:pdf_comparison}), but as the temperature increases, this is no longer the case, justifying the need for molecular dynamics-based sampling.

Finally, the pair distribution functions can be fit to skewed Gaussian distributions (dotted lines in Fig. \ref{fig:pdf_comparison}). Surprisingly, this is sufficient to fully describe all the analyzed pair distribution functions, indicating that the second and third-order interatomic force constants are adequate to provide a complete understanding of the pairwise interactions. This topic will be revisited in Section \ref{sec:thermal_conductivity} when we examine the impact of the fourth-order force constants on thermal conductivity.

\subsection{Quantum effects in the force constants}

With the converged lattice parameters for each temperature of interest, we can now perform (PI)MD simulations under NVT conditions and apply the TDEP method to obtain temperature-dependent second-order IFCs and phonon band structures. 
As the pair-distribution functions in Figure \ref{fig:pdf_comparison} already suggest, we confirm that the differences between phonon dispersions in the classical and quantum sampling cases are minimal at 150K and negligible at higher temperatures. Moreover, no significant changes in the shape of the dispersions are observed at any of the sampled temperatures. This suggests that important components of thermal conductivity, such as phonon frequencies and group velocities, will be largely unaffected by NQEs, which was not expected for very light elements such as boron and nitrogen.

\begin{figure}
    \centering
    \includegraphics[width=\columnwidth]{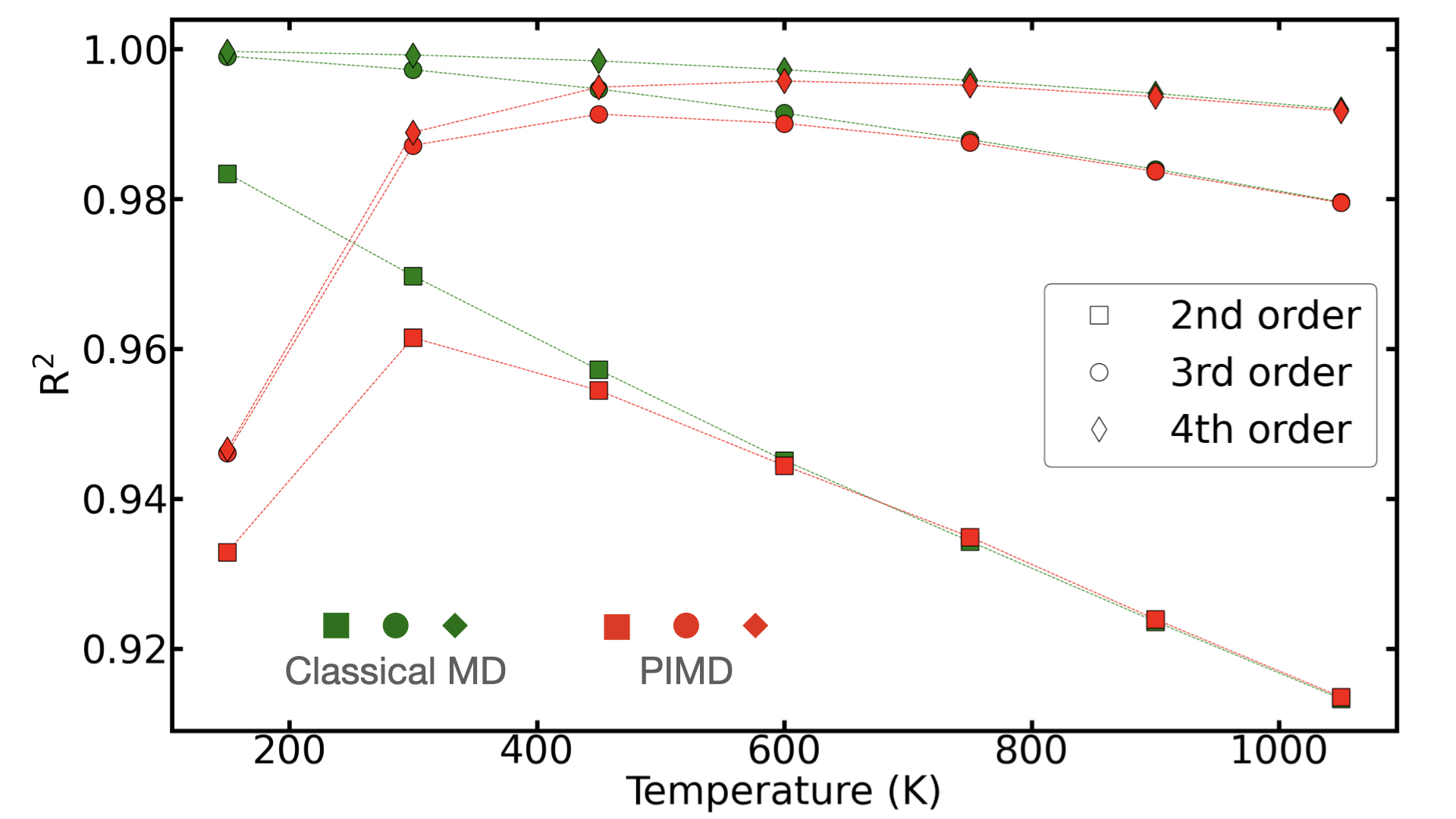}
    \caption{Temperature evolution of the R$^2$ coefficients from the TDEP least-squares fits. In the classical MD sampling case, we observe a decrease in the R$^2$ values with temperature; nonetheless, when fitting up to 4th order the values for R$^2$ are always above 0.99 for all temperature data points. In the PIMD sampling case, the trend is very similar for T $>$ 450 K. For lower temperatures, and especially in the T = 150 K case, the R$^2$ values are significantly lower than their classical counterparts, even including the fit up to 4th order. This is due to the fact that the averaged Born Oppenheimer surface generated in this case is less smooth than in the classical case due to the bead averaging required to construct the centroid dynamics.}
    \label{fig:r2vst}
\end{figure}

\begin{figure}
    \centering
    
    \includegraphics[width=\columnwidth]{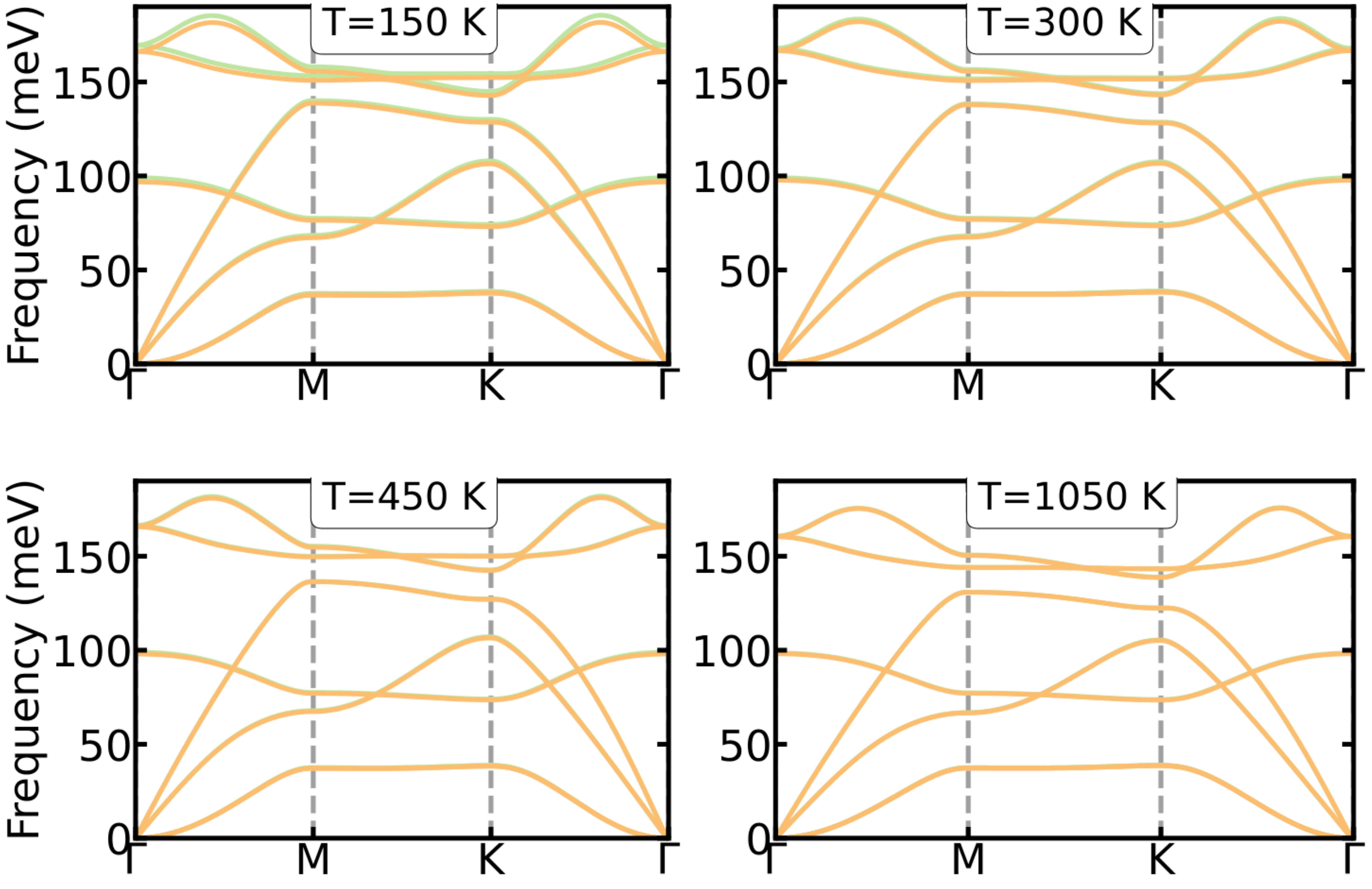}
    
    \caption{Temperature evolution of the phonon band structure. Classical sampling data is presented in green while PIMD sampling data is presented in yellow. We observe negligible changes in all temperatures, with the only noticeable effect being a small vertical shift in the transversal and longitudinal optical modes for T = 150 K. No changes are visible in the shape of the phonon bands due to sampling.}
    \label{fig:phonon_disp}
\end{figure}

Despite the minimal effects on the phonon band structures, the use of PIMD sampling in conjunction with TDEP significantly impacts another metric: the R$^2$ of the force-constant fit, plotted in Fig.~\ref{fig:r2vst}. By obtaining the second, third, and fourth-order IFCs via least square fits, the R$^2$ value indicates how well the forces are described up to the order of the fit. Moreover, this value can indicate the degree of anharmonicity in the system, as a low R$^2$ value of the second-order IFC fit suggests that a significant component of the forces obtained from MD sampling is not harmonic in nature. The anharmonicity measure proposed by Knoop et al. in \cite{Knoop2020} can also be used for the same purpose and is an output of the same routine used to calculate the IFCs in the TDEP code. 

\begin{figure}
    \centering
    \includegraphics[width=\columnwidth]{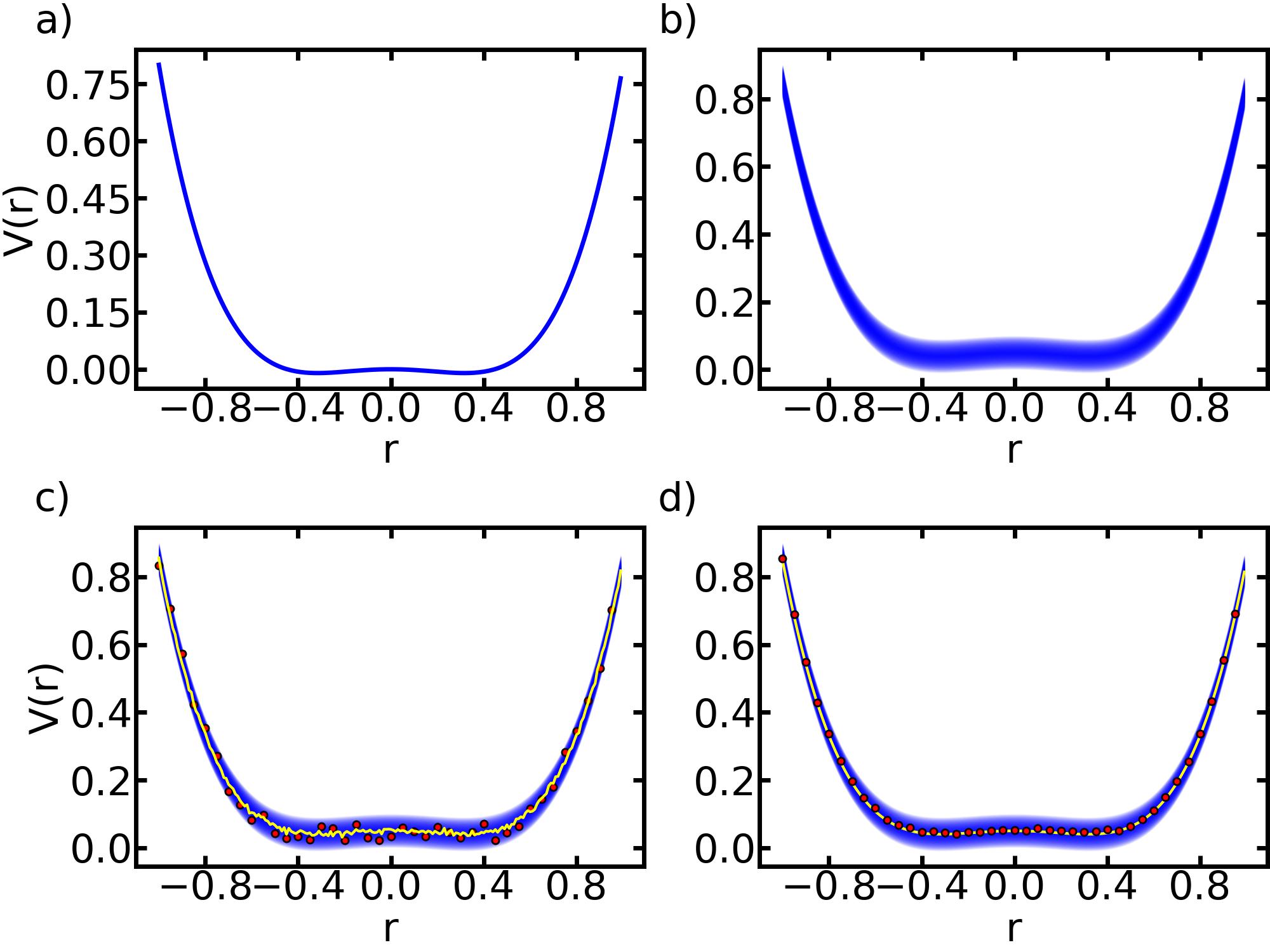}
    \caption{Pictorial representation of the bead averaging effects on the Born-Oppenheimer potential energy surface (PES). Panel a) corresponds to the classical MD case, or equivalently 1 bead, where no averaging happens and the PES is smooth. In figure b) the spread of the quantum ion distribution is shown by the thickness of the color gradient. In c) multiple beads sample the NQE and are averaged to build the centroid. The dots correspond to the averages for each point in the centroid r-grid, with the yellow line being the new PES, now ragged due to noise in the averaging process. Panel d) corresponds to the same case as panel c) but with a very well converged number of beads, where enough data is present to ensure the final PES is again smooth like in panel a)'s case.}
    \label{fig:bhs_pimd}
\end{figure}

However, this is true only in the case of classical MD sampling. An illustrative example can be found in Figure \ref{fig:bhs_pimd}. When performing the TDEP fit with classical MD sampling, the temperature-dependent Born-Oppenheimer surface for the system is indirectly reconstructed. This surface is well-defined as the forces and displacements from a single system's dynamics are used. However, when PIMD sampling is used, a centroid needs to be constructed to approximate the system's Kubo correlation function. In practice, this centroid is constructed by averaging the dynamics of several replica of the physical system, corresponding to the number of beads used for a given temperature case. This averaging process is essential, but the limited number of beads can lead to a noisy description of the effective Born-Oppenheimer surface corresponding to the centroid compared to the smooth one generated in the classical case. This is reflected in the R$^2$ values and can introduce spurious effects, with the only countermeasure being increasing the number of beads for better averaging.

\subsection{Thermal Conductivity} \label{sec:thermal_conductivity}

\begin{figure}
    \centering
    \hspace{-0.5 cm}\includegraphics[width=\columnwidth]{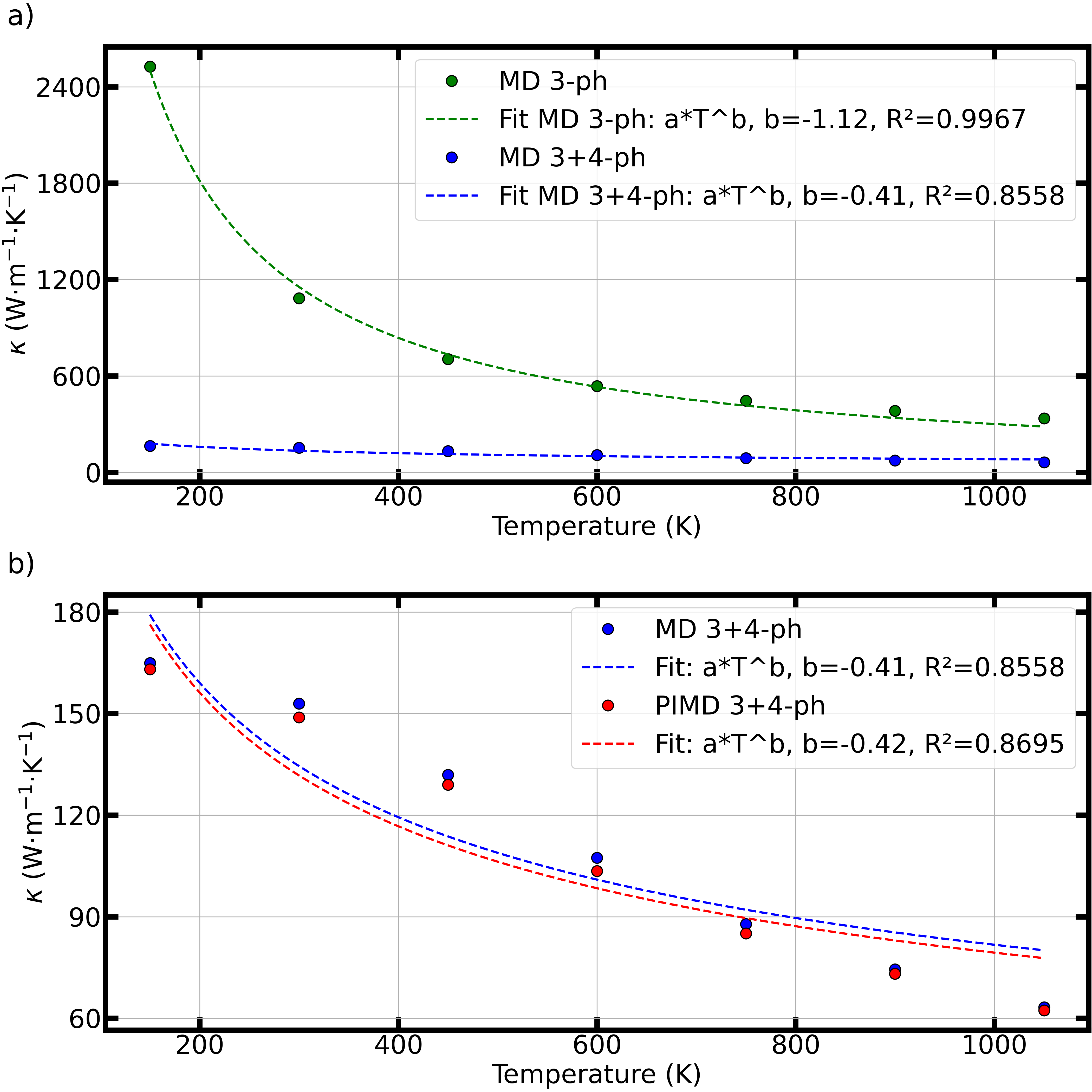}
    \caption{Temperature evolution of the thermal conductivity for monolayer h-BN. Panel a) compares classical MD sampling results using 3-phonon and 3+4-phonon interactions. A drastic reduction is observed when 4-phonon interactions are included, especially at T = 150K where we obtain a reduction of an order of magnitude. In panel b) we compare the classical 3+4-phonon interaction result with it's PIMD counterpart. The results obtained are identical for all temperatures. Isotope scattering is present in all cases as described in the Methods section.}
    \label{fig:compare_kappa}
\end{figure}

\begin{figure}
    \centering

        \centering
        \includegraphics[width=\columnwidth]{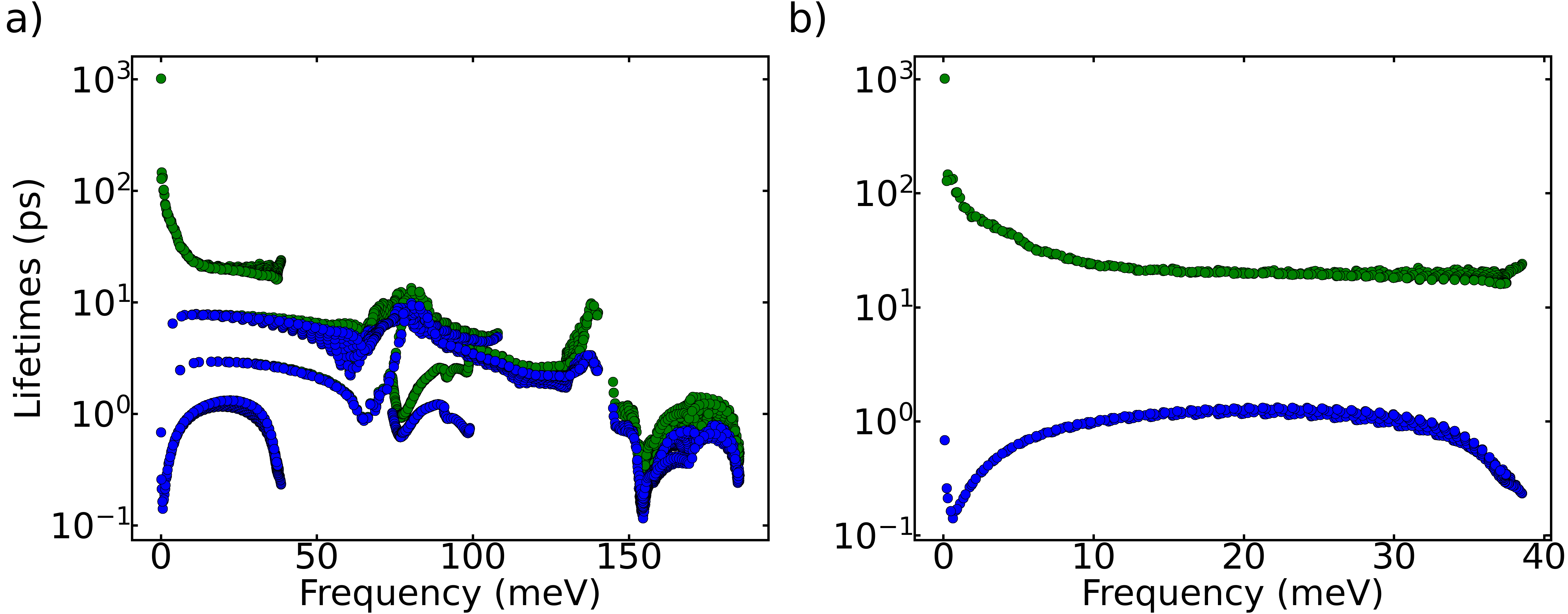}
        \includegraphics[width=\columnwidth]{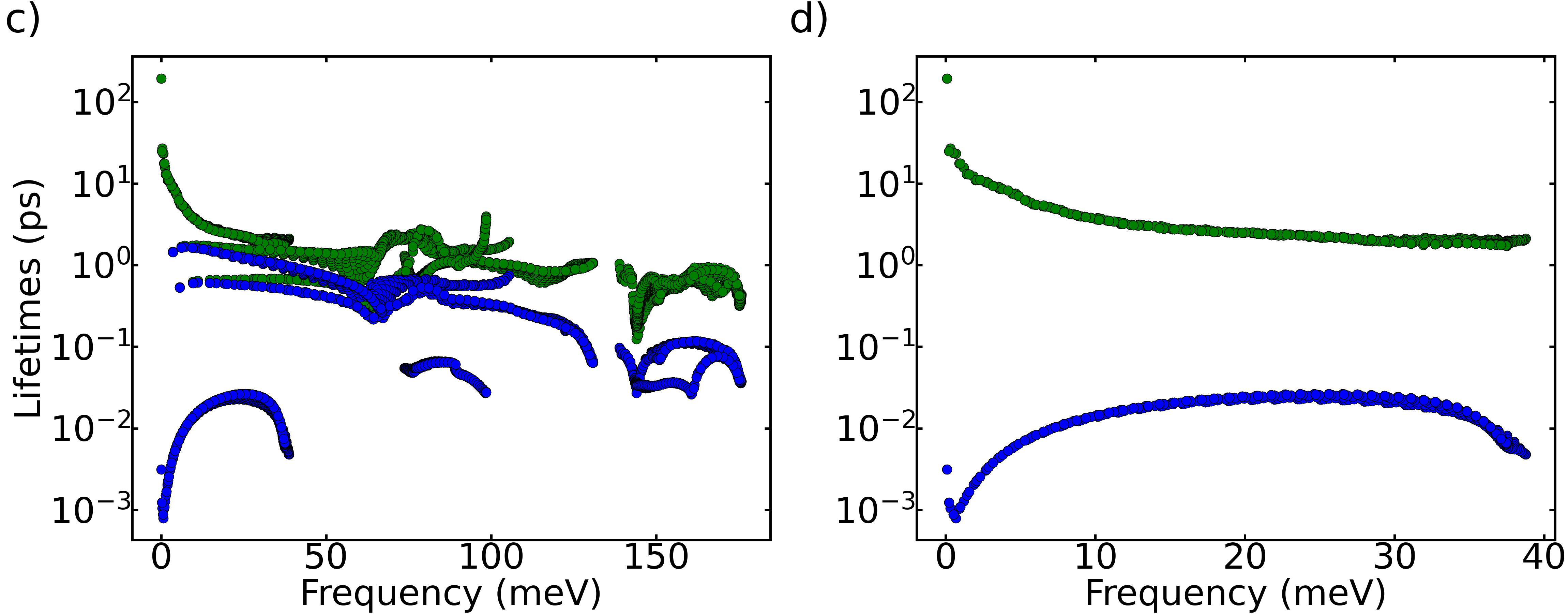}
        
    \caption{Phonon lifetimes for the classical MD sampling case. Figures a) and b) correspond to the T = 150K case for a 60x60x1 q-point grid, while c) and d) correspond to the T = 1050 K case for the same grid. The y-axis is presented in a logarithmic scale. On a) and c) we show the lifetimes for all phonon modes; in b) and d) we single out the acoustic flexural mode, highlighting the difference in magnitude when 4-phonon scattering is included, and the lifetimes are reduced by several orders of magnitude.}
    \label{fig:lifetimes_classical}
\end{figure}

Figure \ref{fig:compare_kappa} summarizes one of the main results of this work. We find that when performing classical MD sampling, the inclusion of 4-phonon scattering results in a substantial reduction of the lattice thermal conductivity for the full temperature range. This effect is particularly noticeable at lower temperatures, where the reduction reaches an order of magnitude for the T = 150 K data point. This result is in agreement with both the results presented in graphene \cite{Han2023,Feng2018,Yukai2023} and other hexagonal boron-X compounds \cite{Feng2016, Feng2017, Kang2018, Bi2022}. The mirror-plane symmetry limits the scattering at the 3-phonon level such that when 4-phonon scattering is included (and thus pairs and quartets of flexural phonons can now be included in interactions) all phonon modes' contributions to the thermal conductivity are reduced.
%
To support the reasoning for this thermal conductivity reduction, we plot the phonon lifetimes for the classical MD case at T = 150 K and T = 1050 K in Figure \ref{fig:lifetimes_classical}. We find that the lifetimes of the acoustic flexural mode are severely reduced (2 orders of magnitude) when 4-phonon interactions are included. This is particularly noticeable at the lowest temperatures, where acoustic modes' contributions to the thermal conductivity are stronger.

\begin{figure}
    \centering
    \includegraphics[width=\columnwidth]{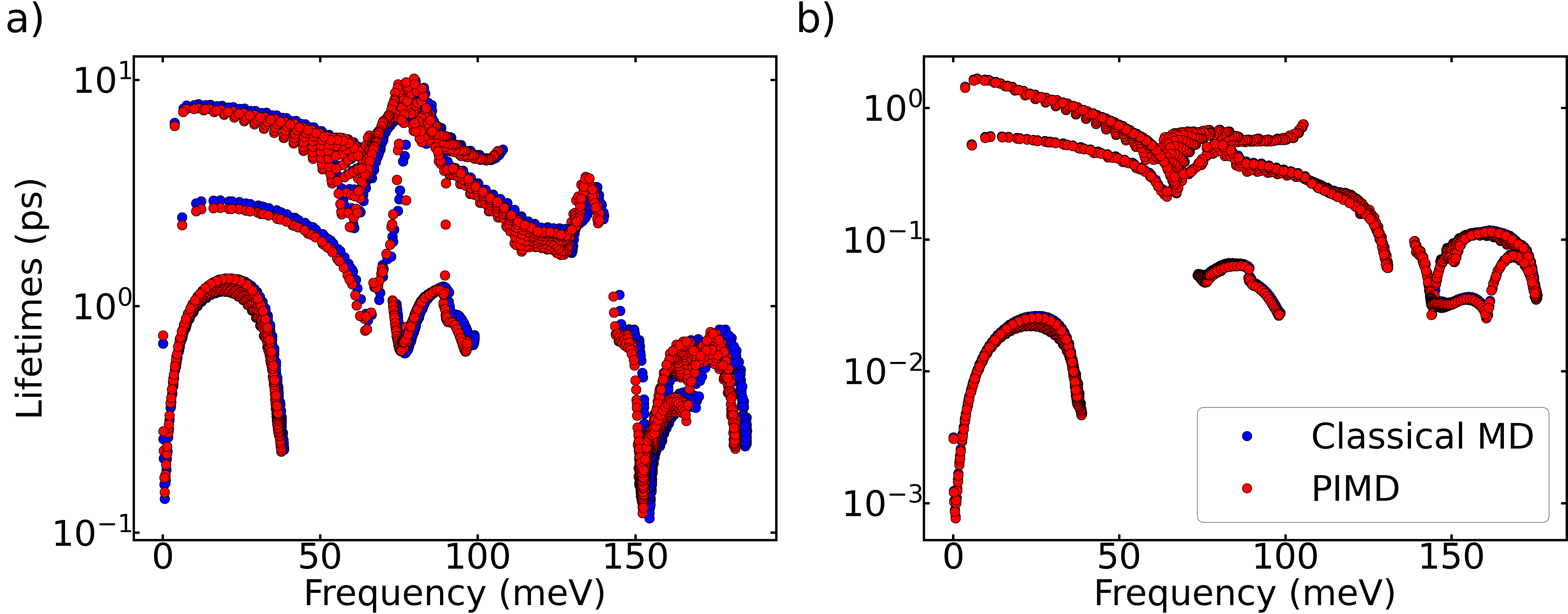}
    \caption{Phonon lifetimes including 3+4-phonon scattering for both classical MD and PIMD sampling for a 60x60x1 q-point grid. Panel a) corresponds to the T = 150 K case while panel b) corresponds to the T = 1050 K case. In both cases we observe that the scattering rates are very close regardless of sampling method, suggesting that quantum effects are negligible.}
    \label{fig:lifetimes_4ph}
\end{figure}

The same trend is observed in panel b) of Fig.~\ref{fig:compare_kappa} when PIMD sampling is performed. Comparing the phonon lifetimes obtained with the classical case in Fig.~\ref{fig:lifetimes_4ph}, we find only very small differences even at the lowest temperatures, and the values obtained for the thermal conductivity match the classical sampling results. We can therefore conclude that, despite the low atomic numbers of both boron and nitrogen, one would have to go far below room temperature to potentially observe nuclear quantum effects in the thermal conductivity of monolayer h-BN. 

We conclude that when computing the lattice thermal conductivity of monolayer h-BN one can safely operate within a classical sampling method without a loss in accuracy. On the other hand, 4-phonon effects are absolutely essential, due to the system's mirror-plane symmetry and to energy conservation restrictions as a consequence of the large gaps in its phonon band structure. It is important to highlight, however, that this need not be the case of other phonon related observables. Indeed as we have shown in Section \ref{sec:sampling} one can fully reproduce the pair-distribution function with great accuracy staying within a 3rd order fit, as the 4th order IFCs are significantly smaller in amplitude than the 3rd order ones. In the thermal conductivity case, however, the suppression of 3rd order processes is such that however small the 4th order IFCs are, they contribute much more to the overall scattering, to the point that for the acoustic flexural mode even the curvature of the lifetimes as function of the frequency changes sign. We therefore expect only observables in which the restriction in scattering phase space is relevant to require 4-phonon scattering to be included. 

It is at this point important to compare our results to previous works. Theoretical approaches to the thermal conductivity of monolayer h-BN can be broadly divided in two groups: BTE based approaches, where phonon IFCs are computed from density functional perturbation theory (DFPT), stochastic sampling, TDEP, etc.; and MD based approaches, such as equilibrium and non-equilibrium molecular dynamics (EMD and NEMD), where the thermal conductivity is directly computed via the Green-Kubo formula \cite{Kubo1957}. In the first case, the usual limitations of previous works consist of the lack of inclusion of isotope and/or 4-phonon scattering. In the second case, while all degrees of anharmonicity are in principle included, isotope effects are neglected. Furthermore, as Tersoff potentials are often used in these cases, results can be very sensitive to their parameterization and on how many nearest-neighbours are included. In both cases, important contributions to the thermal conductivity are systematically neglected, and to our knowledge the first result that included all relevant scattering channels is \cite{Sun2023}, where the room temperature thermal conductivity is reported. 

Regarding EMD based approaches, \cite{Sevik2011} and \cite{Mortazavi2015}
report values of approximately $290$ and $218\ \text{Wm}^{-1}\text{K}^{-1}$ respectively for $\kappa$ at room temperature, while \cite{Khan2017} reports a much larger value of $471\ \text{Wm}^{-1}\text{K}^{-1}$. For NEMD based approaches, \cite{Mortazavi2012} reports a value for $\kappa$ of $58\ \text{Wm}^{-1}\text{K}^{-1}$ also for room temperature. This result is significantly lower than for the EMD based approaches, which the authors attribute to a likely non-optimal parameterization of the Tersoff potential used. Regarding IBTE based approaches, Refs~\cite{Lindsay2011}, \cite{Fan2019} and \cite{Cepellotti2015} report values of $440$, $771$ and $1060\ \text{Wm}^{-1}\text{K}^{-1}$ respectively at room temperature, all of them using DFPT or finite differences to obtain the IFCs and restrict their scattering to isotope and 3-phonon level. More recently, \cite{Sun2023} reported a value for $\kappa$ of $1024.6\ \text{Wm}^{-1}\text{K}^{-1}$ including isotope and 3-phonon scattering, and $229.4\ \text{Wm}^{-1}\text{K}^{-1}$ when 4-phonon scattering is included, using IFCs from a finite difference approach. Their approach is more complete, but the authors do not specify the effective thickness used for the volume normalization of the thermal conductivity, making a direct comparison to our results more delicate. All other values reported above are renormalized with our convention, to reflect the effective thickness previously mentioned in section \ref{sec:methods} and used in all of our own results.

In this work, we obtain within the standard energy conservation approach a value for $\kappa$ at room temperature of $1303\ \text{Wm}^{-1}\text{K}^{-1}$ using only 3-phonon scattering, $1085\ \text{Wm}^{-1}\text{K}^{-1}$ using 3-phonon and isotope scattering, $180\ \text{Wm}^{-1}\text{K}^{-1}$ using 3 and 4-phonon scattering and finally $150\ \text{Wm}^{-1}\text{K}^{-1}$ using isotope, 3 and 4-phonon scattering. 
When comparing to values obtained from EMD approaches, our result using only 3 and 4-phonon scattering is very similar to previously reported values, with differences being easily attributable to the usage of approximate Tersoff potentials. Furthermore, the $20\%$ decrease in $\kappa$ when isotope scattering is included shows an important limitation of these methods. When comparing to IBTE results where isotope and 3-phonon scattering are used, our results match those reported in \cite{Cepellotti2015}. The drastic reduction in $\kappa$ observed in our results when 4-phonon scattering is included shows this level of theory is essential to obtain an accurate, or even qualitatively correct, representation of the thermal conductivity. It is also worth noting that contrary to the previously mentioned results based on the IBTE, in this work temperature-dependent phonon IFCs are included through the TDEP method, while previous works' IFCs are inherently temperature-independent: renormalization is not included when using DFPT, formally at T = 0K. 
Finally, when comparing our results to those reported in \cite{Sun2023}, if one assumes they employed the commonly used effective thickness of 0.33 nm we obtain $745\ \text{Wm}^{-1}\text{K}^{-1}$ with isotope and 3-phonon scattering and $167\ \text{Wm}^{-1}\text{K}^{-1}$ with 4-phonon scattering, which agrees with our results in the latter case but not in the former.

The fact that our result points towards a lower value of $\kappa$ than previously expected and measured experimentally in \cite{Cai2019} and \cite{Ying2019} shows a disconnect exists between experimental and theoretical works. One possible explanation for this difference could be the exposure of the samples to air or other atmospheres other than vacuum. This has been shown \cite{Farris2022, Chen2010} to have an effect on the thermal conductivity, by providing an extra cooling channel to the sample that is not accounted for in the equations used for $\kappa$ and thus leads to a larger apparent value than the intrinsic one obtained when the experiment is done in vacuum. The amount of experimental works done regarding the thermal conductivity of monolayer h-BN is, however, very small and more detailed studies are needed before stronger conclusions can be drawn.

\section{Conclusions}

In summary, the impact of nuclear quantum effects and 4-phonon scattering in the lattice thermal conductivity of monolayer hexagonal boron nitride is investigated using classical molecular dynamics and path-integral molecular dynamics as sampling methods combined with the temperature dependent effective potential approach to compute temperature dependent phonon properties. We show that while nuclear quantum effects are not pronounced for this system, 4-phonon scattering is crucial for a proper description of this quantity as well as its temperature evolution. We further show that these effects are relevant despite not being apparent in the phonon pair distribution functions, as these are well described by skewed Gaussian distributions for all temperatures studied. The consequences of using path-integral molecular dynamics instead of classical molecular dynamics are also discussed, in particular by showcasing its effect on the R$^{2}$ coefficients of the temperature dependent effective potential's least-squares fits and how the bead averaging affects the Born-Oppenheimer potential energy surface, by first making it more ragged for small bead numbers and smoother as the number of beads increases. Finally, we compare our results to previous theoretical and experimental works. In the theoretical works, the lack of either isotope or 4-phonon scattering in the $\kappa$ calculations provides an explanation as to why those values range from 550-650 $\text{Wm}^{-1}\text{K}^{-1}$ at room temperature, instead of our 150 $\text{Wm}^{-1}\text{K}^{-1}$ including both of those effects. Our results point towards a potential disconnect between theoretical and experimental results, and show that further studies are needed on monolayer hexagonal boron nitride to further clarify the origin of their differences.

As an outlook, both graphene and h-BN have particular band structures, where the acoustic and optical band widths are extremely large throughout most of their Brillouin zone. This further restricts the allowed 3-phonon scattering, in a way completely independent from the mirror plane symmetry, by now imposing a constraint based on the energy conservation in the scattering processes\cite{Broido2020, Ravichandran2021}. This second kind of constraint is not exclusive to 2D systems, and has been shown to lead to major effects on the thermal conductivities of boron arsenide and boron phosphide, for example \cite{Feng2016, Feng2017, Kang2018, Bi2022}. When energy conservation conditions are corrected to be compliant with the fluctuation-dissipation condition, however, the impact of this restriction on BAs has been shown to be drastically reduced, and the importance of 4-phonon scattering is greatly diminished when compared to isotope and 3-phonon scattering \cite{Castellano2025b}. The energy conservation restriction, whether fluctuation-dissipation compliant or not, is independent of the mirror-plane symmetry constraint for planar materials.
It will be important in the future to consider the interaction between the 2D effects imposing 4-phonon scattering and the fluctuation dissipation relaxation of the energy conservation.
Work is ongoing in this direction.

\begin{acknowledgments}
The authors acknowledge the Fonds de la Recherche Scientifique (FRS-FNRS Belgium) and Fonds Wetenschappelijk Onderzoek (FWO Belgium) for EOS project CONNECT (G.A. 40007563), and 
F\'ed\'eration Wallonie Bruxelles and ULiege for funding ARC project DREAMS (G.A. 21/25-11).
MJV acknowledges funding by the Dutch Gravitation program
“Materials for the Quantum Age” (QuMat, reg number 024.005.006), financed by the Dutch Ministry of Education, Culture and Science (OCW).

Simulation time was awarded by 
by PRACE on Discoverer at SofiaTech in Bulgaria (optospin project id. 2020225411), 
EuroHPC-JU award EHPC-EXT-2023E02-050 on MareNostrum 5 at Barcelona Supercomputing Center (BSC), Spain
by the CECI (FRS-FNRS Belgium Grant No. 2.5020.11),
and by the Lucia Tier-1 of the F\'ed\'eration Wallonie-Bruxelles (Walloon Region grant agreement No. 1117545).
\end{acknowledgments}

\bibliography{biblio}

\end{document}